\documentclass[%
 reprint,
 amsmath,amssymb,
 aps, prl,
]{revtex4-2}

\usepackage{morefloats}
\usepackage{color}
\usepackage{epsfig,graphicx,amsfonts,amsbsy}
\usepackage{amsmath,amsfonts,amsthm,amssymb}
\usepackage{appendix}
\usepackage{bbm}
\usepackage{makeidx}
\usepackage{url}
\usepackage{verbatim}
\usepackage{mathrsfs} 
\usepackage{morefloats}
\usepackage{comment}
\usepackage{appendix}
\usepackage{bbm}
\usepackage{makeidx}
\usepackage{url}
\usepackage{verbatim}
\usepackage[bookmarksnumbered,pdfpagelabels=true,plainpages=false,colorlinks=true,linkcolor=blue,citecolor=blue,urlcolor=blue]{hyperref}
\usepackage{array}
\usepackage{booktabs}
\usepackage{multirow}
\usepackage{bbm}
\usepackage{tabularx}
\usepackage{cancel,soul}
\usepackage{nicefrac, xfrac}
\usepackage{ulem}
\usepackage{bbold}
\usepackage{comment}
\usepackage{xcolor}

\usepackage[english]{babel}
\usepackage{amsmath}
\usepackage{graphicx}
\def \fcfm {Departamento de F\'isica, CEDENNA, FCFM, Universidad de Chile, Santiago, 8370448, Chile.}
\def \usach {Departamento de F\'isica, CEDENNA, Universidad de Santiago de Chile, Santiago, 9170124, Chile.}

\begin{document}
\title{Electrically Switchable Spintronics in a Multiferroic Altermagnet}
\author{Martin Latorre$^1$, Sebastian Allende$^2$, and Alvaro S. Nunez$^1$}
\affiliation{$^1$\fcfm, $^2$\usach}

\begin{abstract}
We introduce a minimal model of a two-dimensional lattice that, upon spontaneous symmetry breaking, simultaneously develops altermagnetic order, a finite electric polarization, and a spin-polarized transport response, all of which are controlled by an external electric field. By coupling a dimerized altermagnet to an external electric field, we show that the three order parameters are not merely compatible but dynamically entangled, so that switching one (for instance, reversing the polarization with an electric field) necessarily reconfigures the other two. We show that this model offers a clear physical blueprint for designing next-generation spintronic logic and pure spin current memdevices that merge the ultrafast, stray-field-free advantages of compensated magnets with the low-power switching architectures of ferroelectrics.
\end{abstract}

\maketitle

\noindent{\small\textit{This is the accepted manuscript of an article accepted for publication in Physical Review Letters. The published version is available at \href{https://doi.org/10.1103/plh8-5ss1}{https://doi.org/10.1103/plh8-5ss1}. \copyright\ 2026 American Physical Society.}}
\vspace{0.5em}

\paragraph{Introduction.-} Multiferroic materials, in which two or more ferroic orders --- typically ferroelectricity and some form of magnetism --- coexist within a single phase, have been a cornerstone of condensed matter research for the past two decades \cite{Fiebig2005,SpaldinFiebig2005, Urru2020, Rising2021, Saez2023, Mostovoy2024,Lin2025CoulombDrag,Chen2025ElectricalSwitching}. The interest in these systems stems not only from the richness of the underlying physics, where lattice, charge, orbital, and spin degrees of freedom are intimately intertwined, but also from the technological promise of magnetoelectric coupling \cite{Acosta2025}: the possibility of controlling magnetic order with an electric field, or vice versa, without the dissipative costs associated with driving currents through a material \cite{CheongMostovoy2007,Khomskii2009}. The classification of multiferroic mechanisms --- from improper, spin-driven ferroelectricity to lone-pair and charge-ordering routes --- has provided a rich taxonomy that continues to guide the search for new magnetoelectric compounds.

In parallel, the discovery of altermagnetism has reshaped the landscape of unconventional magnetic order. Altermagnets are collinear, compensated magnets with zero net magnetization, like conventional antiferromagnets, but their magnetic sublattices are connected by rotational rather than translational or inversion symmetries \cite{Zeng2026_2DReview}. This distinct spin-group symmetry lifts the spin degeneracy of the electronic bands in momentum space, producing $d$-, $g$-, or $i$-wave spin-split band structures even in the absence of spin-orbit coupling \cite{Smejkal2022a,Smejkal2022b, JungwirthAltermagneticSpintronics, Krempasky2024, Lee2024, Reimers2024, Fedchenko2024, drrnagel2024altermagnetic-a30, gomonay2024structure-3ef,Camerano2025, Meier2026, Wang2025SpinOrbital, pan2025orbital, AntiferroaxialAltermagnetism,TwistedMagnetic}. This nonrelativistic spin splitting endows altermagnets \cite{dong} with phenomenology previously thought to be exclusive to ferromagnets --- such as giant tunneling magnetoresistance, spin currents, and anomalous transport signatures --- while retaining the zero-stray-field, ultrafast spin-dynamics advantages of antiferromagnets \cite{Smejkal2022c}. These properties have positioned altermagnetism as a promising platform for spintronic devices \cite{bai} that combine speed, scalability, and robustness against external fields \cite{JungwirthAltermagneticSpintronics,Zhang2026Staggered}.

The natural convergence of these two research programs --- multiferroics and altermagnetism --- has only recently begun to be explored. Because the altermagnetic order parameter is itself an anisotropic, ferroically ordered object (an even-parity, higher-multipole analogue of a magnetic octupole or similar tensor), it couples naturally to strain and to polar lattice distortions, suggesting that altermagnetism and ferroelectricity need not be independent orders but may instead be cooperatively linked through the same structural degrees of freedom \cite{BhowalSpaldin2024,altermagneticwei, Sun2026}. If such a coupling can be engineered or found in real materials, the resulting compounds would combine electric-field-switchable altermagnetic spin splitting with the low-power, non-volatile control characteristic of ferroelectrics --- a combination that neither conventional multiferroics nor conventional altermagnets can offer on their own.

Despite this promise, a minimal, symmetry-transparent theoretical framework that captures how altermagnetic, ferroelectric, and spin-transport (ferrospintronic\cite{Saez2024}) orders can emerge \emph{simultaneously} from a common microscopic origin is still missing. In this work, we address this gap by introducing a toy model of a two-dimensional lattice that, upon spontaneous symmetry breaking, simultaneously develops altermagnetic order, a finite electric polarization, and a spin-polarized transport response, all controlled by a single set of coupling constants. The model is deliberately minimal: it is built to isolate the symmetry conditions and microscopic ingredients that are sufficient for the three orders to coexist and to be mutually coupled, rather than to describe any specific compound. We show that within this framework the three order parameters are not merely compatible but dynamically entangled, so that switching one (for instance, reversing the polarization with an electric field) necessarily reconfigures the other two.

We believe that this toy model, precisely because of its simplicity, can serve as a template for identifying and engineering real altermagnetic multiferroics, and that it opens an avenue of research at the intersection of unconventional magnetism, ferroelectricity, and spintronics\cite{JungwirthAltermagneticSpintronics, DalDin2024, Grollier2016} that remains largely unexplored. 

\paragraph{Dimerized Altermagnetic Model.-} We base our study on the two-orbital square-lattice model of Leeb et al.~\cite{Leeb2024,DelaBarrera2025bias}. While minimal, this model captures the essential topological features of $d$-wave altermagnets \cite{ma}: a nonrelativistic, momentum-dependent spin splitting of the electronic bands driven by the coexistence of antiferromagnetic and orbital orders. To introduce polar degrees of freedom, we promote this system to a dimerized altermagnet through a Su--Schrieffer--Heeger (SSH) modulation of the nearest-neighbor bonds. The resulting altermagnet is similar to a two-dimensional spin-dependent Rice-Mele Hamiltonian\cite{Rice1982} such as the one in \cite{Nunez2014, Vergara2024, Saez2023, Saez2024, Castro2024}. The full Hamiltonian is
\begin{equation}
    \mathcal{H} = \mathcal{H}_t + \mathcal{H}_J
    + \mathcal{H}_V + \mathcal{H}_{\mathrm{pol}},
    \label{eq:Hfull}
\end{equation}
where $\psi^{\dagger}_{\mathbf{r}\alpha\sigma}$ creates an electron at site
$\mathbf{r}=(i_x,i_y)$ with orbital $\alpha$ in $\{d_{xz},d_{yz}\}$ and spin
$\sigma$ in $\{\uparrow,\downarrow\}$.

We dimerize their kinetic part by modulating the nearest-neighbor
amplitudes over $2\times2$ plaquettes, $t_{1,2}\to t_{1,2}(1\pm\delta)$ on strong
(intra-plaquette) and weak (inter-plaquette) bonds\cite{Vergara2024, Liu2017}, controlled by a single
dimerization $\delta$, while the diagonal amplitudes $t_3$ and
$t_4$---and with them the altermagnetic spin splitting---remain uniform.
Grouping the orbitals into a site spinor
$\Psi_{\mathbf{r}} = (\psi_{\mathbf{r}x\sigma},\psi_{\mathbf{r}y\sigma})^{\mathsf{T}}$,
for which the hopping is diagonal in spin, the kinetic Hamiltonian becomes
\begin{equation}
    \mathcal{H}_t
    = -\sum_{\mathbf{r}}\sum_{\boldsymbol{\rho}}
    \bigl[\Psi^{\dagger}_{\mathbf{r}}\,\mathbf{t}_{\boldsymbol{\rho}}(\mathbf{r})\,
    \Psi_{\mathbf{r}+\boldsymbol{\rho}} + \mathrm{h.c.}\bigr],
    \label{eq:Htdim}
\end{equation}
with $\boldsymbol{\rho}\in\{\hat{x},\hat{y},\hat{x}+\hat{y},\hat{x}-\hat{y}\}$ the
four forward bond vectors to first and second neighbors, the Hermitian
conjugate generating the reverse hops along $-\boldsymbol{\rho}$. The first-neighbor matrices are
diagonal in orbital, $\mathbf{t}_{\hat{x}} = (1\pm\delta)\,\mathrm{diag}(t_1,t_2)$
and $\mathbf{t}_{\hat{y}} = (1\pm\delta)\,\mathrm{diag}(t_2,t_1)$, carrying both
the orbital anisotropy and the dimerization, while the uniform second-neighbor
matrices $\mathbf{t}_{\hat{x}\pm\hat{y}}$ carry $t_3$ on the diagonal
(intra-orbital) and $\mp t_4$ off it (inter-orbital), the sign reversal between
the two diagonal directions being the $d$-wave hallmark of the altermagnet.
Taking the $2\times2$ plaquette as the unit cell folds the Brillouin zone, so
that $\mathcal{H}_t^{\mathrm{dim}}$ is diagonal in the crystal momentum
$\mathbf{K}$ of the reduced zone, with a $16\times16$ Bloch matrix $h(\mathbf{K})$
(four sites, two orbitals, two spins). Strong (intra-plaquette) bonds connect
sites of the same cell and give $\mathbf{K}$-independent entries, whereas weak
(inter-plaquette) bonds link neighboring cells and contribute phases
$e^{\pm iK_{x,y}}$. The uniform limit $\delta\to0$ unfolds the zone and recovers
\cite{Leeb2024}.

The magnetic and orbital interactions are captured, respectively, by a
Heisenberg-like spin exchange and an Ising-like orbital exchange,

\begin{equation}
    \mathcal{H}_J = J\!\sum_{\langle\mathbf{r}\mathbf{r}'\rangle}
    \mathbf{S}_{\mathbf{r}}\!\cdot\!\mathbf{S}_{\mathbf{r}'},
    \quad
    \mathcal{H}_V = V\!\sum_{\langle\mathbf{r}\mathbf{r}'\rangle}
    N^{z}_{\mathbf{r}}N^{z}_{\mathbf{r}'},
    \label{eq:HJHV}
\end{equation}

where
$\mathbf{S}_{\mathbf{r}} = \sum_{\alpha\sigma\sigma'}\psi^{\dagger}_{\mathbf{r}\alpha\sigma}
\tfrac12\boldsymbol{\sigma}_{\sigma\sigma'}\psi_{\mathbf{r}\alpha\sigma'}$ is the local
spin operator and
$N^{z}_{\mathbf{r}} = \sum_{\alpha\beta\sigma}\psi^{\dagger}_{\mathbf{r}\alpha\sigma}
\tau^{z}_{\alpha\beta}\psi_{\mathbf{r}\beta\sigma}$ the orbital pseudospin operator, with
$\boldsymbol{\sigma}$ and $\tau^{z}$ the Pauli matrices acting on spin indices
$\sigma,\sigma'$ in $\{\uparrow,\downarrow\}$ and orbital indices
$\alpha,\beta$ in $\{d_{xz},d_{yz}\}$, respectively. In the altermagnetic region
$J  > 0$ drives $(\pi,\pi)$ antiferromagnetic order and
$V > 0$ drives $(\pi,\pi)$ orbital order; together these two
orders define the altermagnetic phase.

\paragraph{Coupling to an external electric field.-} 
The final term $\mathcal{H}_{\mathrm{pol}}$ is a staggered on-site potential that breaks the inversion symmetry of the lattice,
\begin{equation}
    \mathcal{H}_{\mathrm{pol}} = e\,a\,E_y\sum_{\mathbf{r}}\xi_{\mathbf{r}} \sum_{\alpha\sigma}\psi^{\dagger}_{\mathbf{r}\alpha\sigma}\psi_{\mathbf{r}\alpha\sigma},
    \label{eq:Hpol}
\end{equation}
where $\xi_{\mathbf{r}}=(-1)^{i_y}$ is a $(0,\pi)$ sublattice modulation, chosen odd under the plaquette-center inversion $\mathcal{P}$. Because this potential is independent of spin and orbital, any resulting spin polarization is inherited exclusively from the altermagnetic band structure.

Rather than a fixed ionic potential, the amplitude is controlled by an external electric field $\mathbf{E}$ applied to a rigid lattice. Projecting the scalar potential $\phi(\mathbf{r})=-E_y y$ onto the
$2\times 2$ cell separates it into a cell-average part, which grows linearly
from cell to cell and carries the non-periodicity, and an intracell staggered
component $\propto\xi_r$, which is the same in every cell.
Retaining only the latter is what makes the problem periodic; the discarded
ramp is what drives Zener tunneling and Wannier--Stark localization, which we
neglect in the quasi-static regime considered here. This explicitly captures the coupling $-\mathbf{p}\cdot\mathbf{E}$ of the field to the intracell dipole $\mathbf{p}\parallel\hat{y}$ formed by the two rows of the plaquette separated by $a$. This reduction maps the uniform field to a well-defined periodic Rice--Mele mass $e\,a\,E_y$, bypassing the gauge ambiguities of a scalar potential in a periodic crystal. The dimerization preserves the defining symmetry of the
altermagnetic state: because $t_{\hat x}$ and $t_{\hat y}$ carry the same bond
modulation with the two orbital roles interchanged, the plaquette-centred
rotation combined with the spin flip, $[C_2\|C_{4z}]$, remains an exact
symmetry at $E=0$, and the model stays a $d$-wave
altermagnet~\cite{Smejkal2022a}. The same operation maps $\xi_r$ onto the
corresponding profile along $\hat x$, so that $E_x$ and $E_y$ are equivalent
up to a reversal of the sign of $P_S$; the choice of $\hat y$ is a
convention.

\begin{figure}[h!]
    \centering
    \includegraphics[width=1\columnwidth]{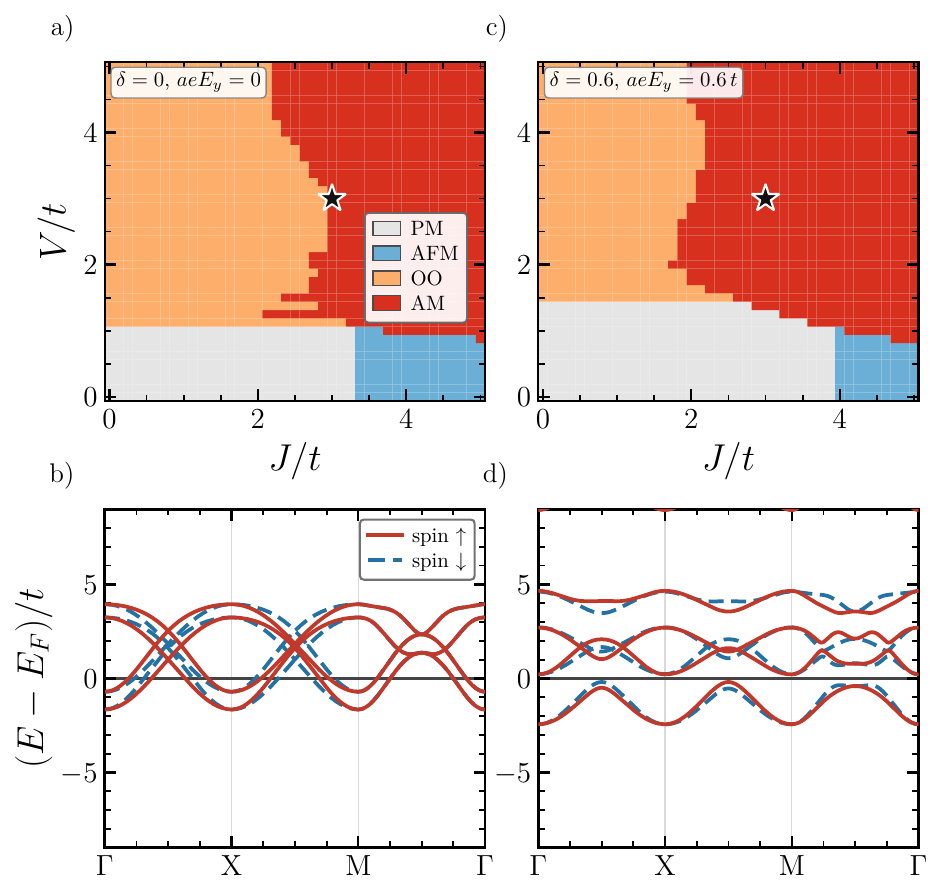}
    \caption{%
    Ground-state phase diagram and spin-resolved band structure of the dimerized
    altermagnet, in units of the master hopping $t\equiv|t_1|$ at one-eighth filling.
    (\textbf{a,c})~Mean-field phase diagram in the $J$--$V$ plane, with the four phases
    (PM, AFM, OO, AM) read from the staggered amplitudes $(m,o)$; the star marks the working point $J=V=3t$.
    (\textbf{b,d})~Spin-resolved bands along $\Gamma$--$X$--$M$--$\Gamma$, as $(E-E_F)/t$,
    with spin $\uparrow$ (solid red) and $\downarrow$ (dashed blue) split by the
    $d$-wave altermagnetic mechanism.
    Left column (\textbf{a,b}): undimerized reference, $\delta=0$, $eaE_y=0$. Right column
    (\textbf{c,d}): $\delta=0.6$, $eaE_y=0.6\,t$, where the SSH modulation opens gaps
    while leaving the splitting intact.
    \label{fig:combined}}
\end{figure}

This field-induced potential is the two-dimensional generalization of the Rice--Mele mass, analogous to the topological multiferroic constructed in Ref.~\cite{Vergara2024}. Since the kinetic $\mathcal{H}_t^{\mathrm{dim}}$ and exchange $\mathcal{H}_J,\mathcal{H}_V$ terms are even under $\mathcal{P}$, only $\mathcal{H}_{\mathrm{pol}}$ breaks parity. A finite $E_y$ thus lifts the bulk quantization, enabling a continuous charge polarization and, crucially, a finite, electrically switchable spin polarization $P_s$. It is this field-controlled inversion breaking that endows the system with its multiferroic polarization--spin coupling~\cite{Vergara2024}. We refer to the full $\mathcal{H}$ in Eq.~\eqref{eq:Hfull} as the dimerized altermagnetic model.
\paragraph{Results.-}
\underline{Ground-state phase diagram.}
We first map the mean-field ground state of the dimerized altermagnet in the
plane of the two competing interactions, the Heisenberg spin exchange $J$ and
the Ising orbital exchange $V$, both measured in units of the master hopping
$t\equiv|t_1|$. For each $(J,V)$ we solve the mean-field self-consistency at fixed filling of one eighth and read off the
two staggered order parameters, the N\'eel amplitude
$m$ and the orbital amplitude $o$. Their joint pattern partitions the
plane into four phases: a paramagnetic metal ($m=o=0$), a pure
antiferromagnet ($m>0$, $o=0$), a pure orbital-ordered state ($m=0$, $o>0$),
and the altermagnet ($m>0$ and $o>0$), in which $(\pi,\pi)$ N\'eel and
$(\pi,\pi)$ orbital order coexist~\cite{Leeb2024,DelaBarrera2025bias}
[Fig.~\ref{fig:combined}(a,c)].
The topology of the diagram follows directly from the microscopics rather than
from any imposed constraint. Along $J=0$ the spin channel cannot order and the system passes from paramagnet to orbital-ordered state with increasing $V$; the orbital channel condenses first, so the orbital-ordered lobe dominates the diagram while the altermagnet occupies the adjacent large-$J$, large-$V$ region. The altermagnet
contains the working point $J=V=3t$, where the order is well developed
($m=0.08$, $o=0.48$ at $\delta=0$, the orbital amplitude setting the dominant scale). This is the regime relevant to a
well-ordered altermagnetic insulator and the reference point for all subsequent
results.
The key structural result is how this phase boundary responds to the two symmetry-breaking fields of our model: bond dimerization $\delta$ and the field-induced polar mass $aeE_y$ [Fig.~\ref{fig:combined}(a) vs.~(c)]. Each enlarges the altermagnetic region. With dimerization ($\delta=0.6$) and polar field ($eaE_y=0.6\,t$), the altermagnetic area grows from $38\%$ to $45\%$ of the plane, advancing the onset of altermagnetism to weaker exchange. The mechanism is clear: the SSH modulation and the Rice--Mele mass open a single-particle gap that removes the competing metallic screening, so that a N\'eel amplitude is stabilized on top of the pre-existing orbital order ($m$ rising from $0.08$ to $\approx0.23$ at the working point). Crucially, the altermagnet is not destroyed but reinforced. The working point remains deep inside the altermagnetic phase for all $\delta$ and $E_y$ studied, showing that dimerization and electric field promote, rather than antagonize, altermagnetic order. This is the prerequisite for using the field as a nonvolatile control
knob without quenching the magnetic state it is meant to read out.

\underline{Spin-resolved band structure.} The defining fingerprint of altermagnetism is a spin splitting of the
electronic bands in the complete absence of spin--orbit coupling and net
magnetization~\cite{Smejkal2022a,Smejkal2022b}. Because our Hamiltonian conserves $S^z$, it is block-diagonal in spin, and the splitting
$\Delta_{\mathbf k}^{\sigma}=\varepsilon_{\uparrow}(\mathbf k)-
\varepsilon_{\downarrow}(\mathbf k)$ is well defined along the
entire Brillouin-zone path [Fig.~\ref{fig:combined}(b,d)]. Dispersions are plotted as $(E-E_F)/t$, so that occupied and empty states are
separated by zero and the two columns share a common scale
[Fig.~1(b) vs.~(d)].

Orbital order coexists with and defines the altermagnetic phase, but the
spin splitting itself originates from the anisotropic (direction-dependent)
hopping of the parent altermagnet~\cite{Leeb2024}. Bond dimerization
modulates a different set of hoppings and therefore leaves the $d$-wave
splitting essentially intact [Fig.~\ref{fig:combined}(b)~vs.~(d)]; $\delta$
and $E_y$ mainly open and reshape gaps around the Fermi level, rather than
remove the altermagnetic splitting.

\underline{Inversion symmetry and the route to electrical control.} The two figures fix the symmetry setting used in the rest of the paper. Without the polar field, the four ground-state terms---the dimerized kinetic Hamiltonian~\eqref{eq:Htdim} and the two exchange fields~\eqref{eq:HJHV}---are all even under plaquette-center inversion $\mathcal P$, so
$\mathcal P\,h(\mathbf K)\,\mathcal P^{-1}=h(-\mathbf K)$ at $aeE_y=0$. The system is then centrosymmetric in every ordered phase, and the electronic polarization (the $2$D Berry phase of the occupied bands) is quantized; in particular, the spin-resolved polarization $P_S$ is fixed at zero. The polar potential $\mathcal H_{\mathrm{pol}}$~\eqref{eq:Hpol} is the only on-site term odd under $\mathcal P$, and turning it on (a finite applied field $E_y\neq0$) breaks centrosymmetry, lifting the quantization of the $2$D Berry phase and, as we show next, generating a finite, electrically switchable spin polarization in a band structure where the altermagnetic splitting is already present. The phase diagram ensures this switching occurs within the altermagnetic phase, and the band structure ensures a spin-split Fermi sea for the field to act on. As anticipated, the one-eighth-filled ground state is a compensated altermagnetic insulator. Because the bond dimerization $\delta$ preserves spatial inversion, this unperturbed state lacks a spontaneous polar moment, making the external electric field strictly necessary to break centrosymmetry and induce ferroelectricity.
The compensation survives at finite field: since $S^z$ is
conserved and the spectrum remains gapped throughout the sweep, the field
cannot transfer occupation between the two spin sectors, and the net
magnetization $M=\sum_\sigma\sigma N_\sigma$ stays identically zero. The finite $P_S$ is a Berry-phase property of the occupied bands, not a net
moment.

Ferroelectricity is governed by the electric field $E_y$, which breaks inversion symmetry. The free energy as a function of $E_y$ forms a double well with degenerate minima, representing the two states of spontaneous polarization. The system is thus a proper, field-switchable ferroelectric. To quantify this multiferroic response, we compute the macroscopic polarizations using the geometric Berry phase approach\cite{Onoda2004, Vanderbilt2018}: the charge polarization $P_e$ is evaluated via the standard 2D Berry phase of the occupied Bloch bands, while the spin ($P_s$) polarization is obtained analogously by weighting the trace of the non-Abelian Berry connection with the corresponding Pauli matrices ($\sigma^z$) prior to momentum integration.

\begin{figure}[h!]
  \centering
  \includegraphics[width=\linewidth]{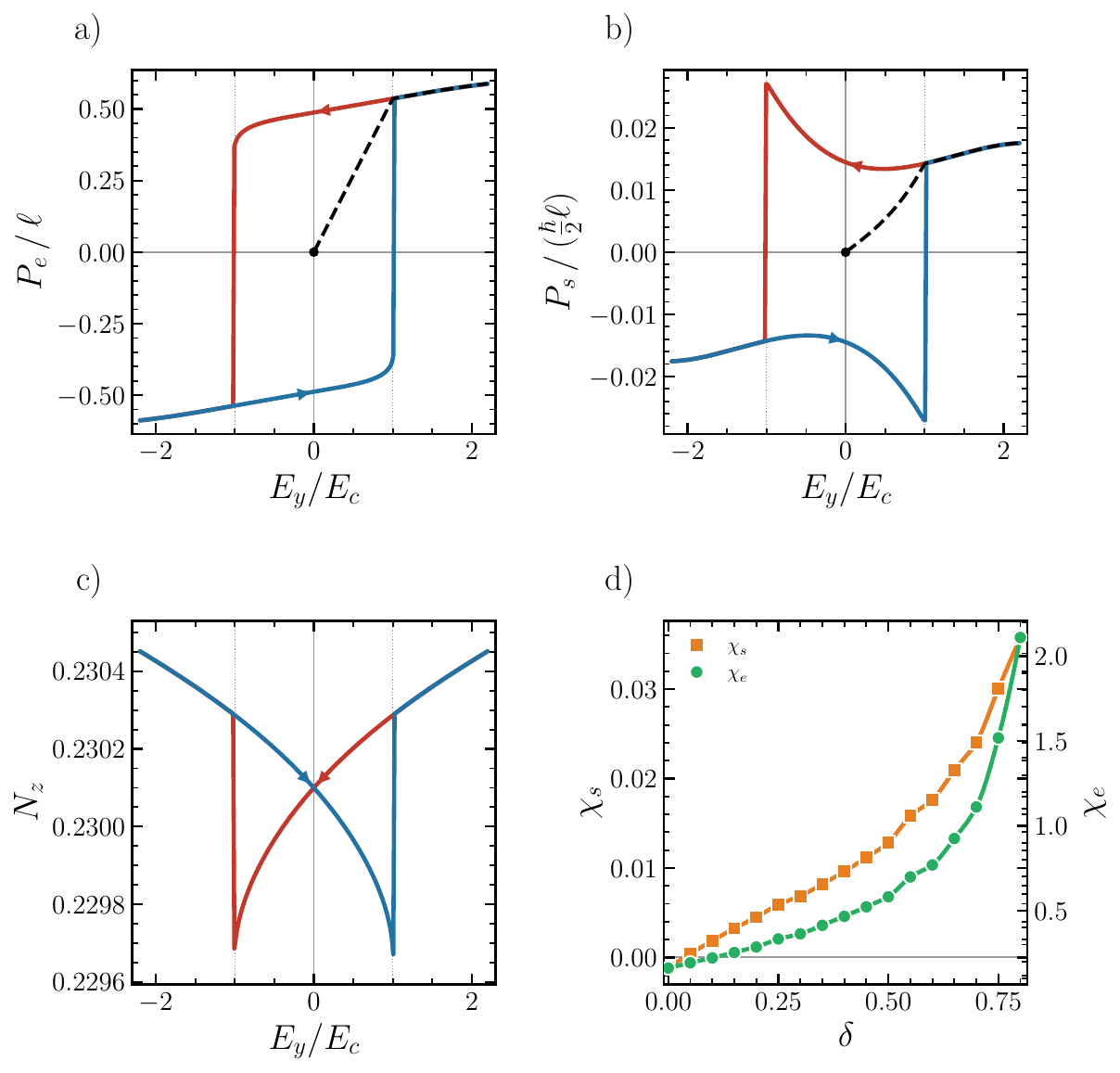}
  \caption{Quasi-static loops of the two polarizations, magnetic order versus the applied electric field $E_y$ and Zero-field multiferroic susceptibility of both polarization
  channels as a function of the dimerization $\delta$ in the altermagnetic state at one-eighth filling ($J=V=3t$, dimerization $\delta=0.6$). Red and blue curves show the two sweep directions; dashed lines mark the virgin curve. (\textbf{a--b}) Charge $P_e$ in units of $\ell = e\,a/A_\mathrm{cell}$ and spin $P_s$ polarizations in units of $\ell \hbar/2$. Both form square, odd-in-$E_y$ loops and switch simultaneously at the same coercive field. (\textbf{c}) The z component of the N\'eel vector $N_z$, even in $E_y$, forming a butterfly loop with maxima at the coercive fields, evidencing magnetoelectric coupling. (\textbf{d}) Each
  $\chi_i=\partial P_i/\partial E_y\big|_{E_y\to0}$ is the initial slope of the
  corresponding quasi-static loop, in normalized units: charge
  ($\chi_e$, green circles) and spin ($\chi_s$, orange squares).}
  \label{fig:loops}
\end{figure}

Sweeping the electric field quasi-statically, the polarization traces a square hysteresis loop\cite{Ma2018} (Fig.~\ref{fig:loops}). At each field step the self-consistency is initialized with the
converged solution of the previous step, so the calculation follows the
locally stable branch rather than the global minimum; the coercive field is
the point at which that branch ceases to exist. The loops therefore reflect
the metastability of the mean-field landscape and do not describe the
switching dynamics, which would involve nucleation and domain-wall motion. The key result is that one polar mode synchronously switches two polarization channels: charge ($P_e$) and spin ($P_s$). Both are odd in the polar mode, vanish in the non-polar state, and jump at the same coercive field, $|eaE_y|/t\simeq 0.034$ (Fig.~2a--b). These memory-like effects might prove valuable for potential applications in memdevices\cite{Chanthbouala2012,CheongMostovoy2007, Tetzlaff2013, Thomas2014, Chua2019}.

\underline{Multiferroic susceptibility.}
Having established that a single polar mode switches both polarization
channels, we quantify how strongly each channel responds to the field in the
linear regime and how that response is controlled by the dimerization.

For each
$\delta$ we extract the zero-field susceptibility of every ferroic channel,
$\chi_i=\partial P_i/\partial E_y\big|_{E_y\to0}$ ($i=e,s$), i.e.\ the initial
slope of the corresponding quasi-static loop (Fig.~\ref{fig:loops}d). This is the
natural linear-response measure of the multiferroic coupling: it records the
charge and spin response induced per unit field before any switching
occurs, in the respective normalized units as the loops, and its
$\delta$-dependence shows how bond dimerization tunes the magnetoelectric
strength.

It is worth distinguishing the two field regimes at this point.
All the hysteretic quantities discussed so far correspond to the
\emph{quasi-static} limit, in which the field is swept slowly and the
non-volatile remanent states are the quantity of interest, whereas the
spin-current generation described next requires a \emph{time-dependent} field
and is therefore a distinct, dynamic effect.
The implications of this finite spin susceptibility extend well beyond the 
static magnetoelectric response, establishing the physical foundation for 
dynamic ferrospintronics. Because the spin polarization is rigidly locked 
to the polar lattice mode, any temporal modulation of the applied electric 
field---such as a high-frequency AC drive or a rapid ferroelectric switching 
event---must be accompanied by a dynamic reconfiguration of the spin 
texture. This symmetry-allowed cross-coupling implies that driving the 
system electrically will intrinsically pump  spin currents through the 
bulk\cite{Shi2006, Nunez2014, Ulloa2017, Vergara2024, Chen2025, Castro2026}. Crucially, this generation mechanism relies entirely on the altermagnetic 
splitting and the inversion-breaking field, completely bypassing the need for 
spin-orbit coupling, macroscopic magnetic moments, or dissipative charge bias. 
Consequently, the dimerized altermagnet operates not merely as a switchable 
magnetic state, but as an active spin-current generator, explicitly linking 
the structural dimerization $\delta$ to the magnitude of the spintronic output.
\begin{figure}[t]
    \centering
    \includegraphics[width=0.85\columnwidth]{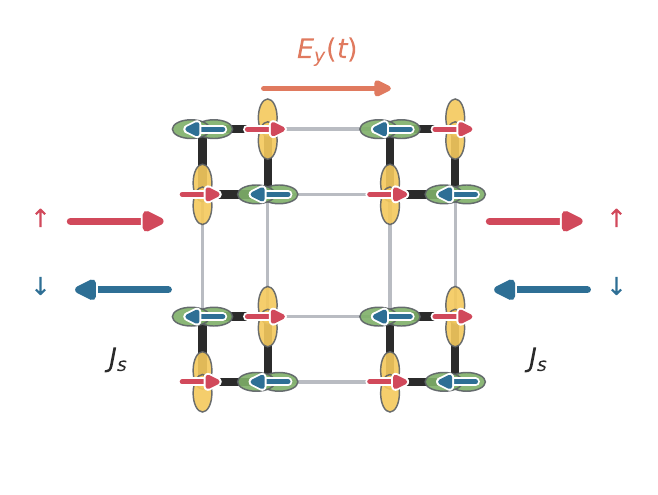}
    \caption{Spin-current generation under a time-dependent field
$E_y(t)$. Red (blue) arrows denote the up (down) spin channel on the two
dimerized sublattices; opposite orbital orientations encode the
altermagnetic $d$-wave order. A time-varying field pumps a
longitudinal spin current, $J_s^{\mathrm{inj}}=Z_s\,\chi_s\,
\mathrm{d}E_y/\mathrm{d}t$, where $\chi_s=\partial P_S/\partial
E_y|_{E_y\to 0}$ is the zero-field spin susceptibility of Fig.~2(d) and
$Z_s\le 1$ is a dimensionless interface injection efficiency. The bulk
response $\chi_s\,\mathrm{d}E_y/\mathrm{d}t=\mathrm{d}P_S/\mathrm{d}t$ is a
displacement current of the ground state, well defined in the gapped state,
and distinct from a Fermi-surface transport current.}
    \label{fig:spin_current}
\end{figure}

\paragraph{Discussion.-} The multiferroic character also appears in the
direct magnetoelectric coupling, seen in the magnitude of the N\'eel vector
(Fig.~\ref{fig:loops}c). Unlike the polarizations, $m$ is an even function of
the polar mode: instead of changing sign it forms a shallow butterfly-shaped
loop, with extrema at the coercive fields---where the system crosses the polar
transition---and a reversible modulation of about $\sim\!0.4\%$, so each
ferroelectric switching event reproducibly changes the amplitude of the
N\'eel order. This opposite parity---polarizations odd, magnetic order
even---follows from the symmetries of the altermagnetic state and
distinguishes this case from a conventional multiferroic.

The even parity fixes the form of the coupling. The linear magnetoelectric
response vanishes by symmetry,
$\partial m/\partial E_y|_{E_y=0}=0$\cite{Castro2025,Eerenstein2006, Wang2016, Visakh2021, Cano2021}, and the
leading term is biquadratic: in a Landau description
$F_\mathrm{ME}=b\,m^{2}(eaE_y)^{2}$, the generic coupling of type-I
multiferroics. Minimizing at fixed field gives
$m(E_y)=m_{0}\,[\,1+g\,(eaE_y/t)^{2}\,]$, which our data follow with
$m_{0}\simeq0.23$ and $g\simeq+0.02$. The positive sign means the polar
distortion weakly reinforces the N\'eel order, so the N\'eel amplitude is
largest in the saturated (remanent) state.
 
\underline{Material relevance.-} These features place our model as the minimal,
fully solvable description underlying the recent surge of altermagnetic
multiferroics. Symmetry analyses and first-principles studies have established
that altermagnetism \cite{jungwirth2026symmetry} and ferroelectricity can coexist and couple: an
altermagnetoelectric cross-coupling between spin and electric polarization has
been proposed in polar magnets such as BaCuF$_4$, Ca$_3$Mn$_2$O$_7$ and
BiFeO$_3$~\cite{Smejkal2024_AME, SeleznevDaniel}, and electric-field switching of the
altermagnetic spin splitting has been demonstrated in the hybrid-improper
ferroelectric [C(NH$_2$)$_3$]Cr(HCOO)$_3$~\cite{Gu2025_FSA}. Most directly, a
family of two-dimensional ferroelectric altermagnets---monolayer vanadium
oxyhalides and sulfide halides (VOX$_2$, VSX$_2$)---realizes the same mechanism, with a Peierls-like bond
dimerization breaking the sublattice equivalence so that reversing the
electric polarization interchanges the sublattice hoppings and flips the spin
polarization~\cite{Zhu2025_2DFEAM}, there with the sublattices related by a mirror
rather than by the fourfold rotation of our lattice. Our
dimerized altermagnet captures this hopping-exchange mechanism in its simplest form, and resolves it to the end: the full quasi-static hysteresis of the charge,
and spin polarizations, the biquadratic modulation of the N\'eel order,
and the spin--electrical locking are quantitative, material-independent predictions
that these candidate compounds should display, and that the proposed
magneto-optical probes can test.

\paragraph{Conclusions.-} In conclusion, we have established a minimal theoretical framework demonstrating the fundamental compatibility and dynamical entanglement of altermagnetism, ferroelectricity, and ferrospintronic order. By coupling a dimerized altermagnet to an external electric field, we showed that a single polar mode synchronously switches the charge and spin dipolar densities while modulating the collinear N\'eel vector through a biquadratic magnetoelectric interaction. 

Crucially, the robust spin--electric locking in this model enables a direct dynamic cross-response: the application of a uniform electric field not only reverses the static polarizations but inherently drives spin-polarized transport, as illustrated in Fig.~\ref{fig:spin_current}. 
This capability to generate and control spin currents purely through an electric field---without relying on spin--orbit coupling, macroscopic magnetization, or external magnetic fields---highlights the immense technological potential of altermagnetic multiferroics. A key payoff is experimental: the theory predicts spin currents induced directly by electric fields, providing clear, testable signatures of the proposed mechanism in transport or optical measurements and opening a realistic route to electrically controlled spin-based devices built on robust altermagnets.
By providing a non-volatile, history-dependent mechanism for spin control, our model offers a clear physical blueprint for designing next-generation spintronic logic and  spin current memdevices that merge the ultrafast, stray-field-free advantages of compensated magnets with the low-power switching architectures of ferroelectrics.

\paragraph{Acknowledgments} This work was funded by ANID CEDENNA CIA 250002.  Funding is acknowledged from Fondecyt Regular 1230515. S.A. acknowledges funding from Fondecyt Regular 1261323.

\bibliography{altermagnetic}

\end{document}